\documentclass[a4paper,preprint,aps]{revtex4}

\usepackage{graphicx}% Include figure files

\newcommand{\eqa}{\begin{equation}}
\newcommand{\eqz}{\end{equation}}
\newcommand{\eqma}{\begin{eqnarray}}
\newcommand{\eqmz}{\end{eqnarray}}

\begin{document}
\title{Heats of formation of perchloric acid, HClO$_4$, and perchloric anhydride, Cl$_2$O$_7$. Probing
the limits of W1 and W2 theory\thanks{For WATOC05 special issue}}
\author{Jan M. L. Martin}
\email{comartin@wicc.weizmann.ac.il} \affiliation{Department of
Organic Chemistry, Weizmann Institute of Science, IL-76100
Re\d{h}ovot, Israel}
\date{{\em J. Mol. Struct.} (WATOC'05 issue); Received June 26, 2005; Revised \today}
\smallskip
\begin{abstract}
The heats of formation of HClO$_4$ and Cl$_2$O$_7$ have 
been determined to chemical
accuracy 
for the first time 
by means of W1 and W2 theory. These molecules exhibit particularly severe degrees of inner polarization, and as such obtaining a basis-set limit SCF component to the total atomization energy 
becomes a challenge. (Adding high-exponent $d$ functions to a standard $spd$ basis set has an effect
on the order of 100 kcal/mol for Cl$_2$O$_7$.) Wilson's aug-cc-pV(n+d)Z basis sets represent a dramatic
improvement over the standard aug-cc-pVnZ basis sets, while the aug-cc-pVnZ+2d1f sequence converges still
more rapidly. Jensen's polarization consistent basis sets still require additional high-exponent $d$
functions: for smooth convergence we suggest the \{aug-pc1+3d,aug-pc2+2d,aug-pc3+d,aug-pc4\} sequence.
The role of the tight $d$ functions is shown to be an improved description of the Cl (3d) Rydberg orbital, enhancing its ability to receive back-bonding from the oxygen lone pairs. In problematic cases like this (or indeed in general),
a single 
SCF/aug-cc-pV6Z+2d1f calculation
may be preferable over empirically motivated extrapolations.
Our best estimate heats of
formation are $\Delta H^\circ_{f,298}[$HClO$_4$(g)$]=-0.6\pm$1 kcal/mol and 
$\Delta H^\circ_{f,298}[$Cl$_2$O$_7$(g)$]=65.9\pm$2 kcal/mol, the largest source of uncertainty being our 
inability to account for post-CCSD(T) correlation effects. While G2 and G3 theory have fairly large errors,
G3X theory reproduces both values to within 2 kcal/mol.
\end{abstract}
\maketitle

\section{Introduction}

The author's Dirac medal lecture at WATOC$'$05 focused on some recent advances in the area of computational thermochemistry in general, and on the W$n$ (Weizmann-$n$) theory\cite{w1,w1eval,w1review,w3} developed in our own laboratory in particular. The subject has been reviewed very recently\cite{compthermoreview} and therefore will not be re-reviewed here.
Rather, we will present an example practical application.

Perchloric acid, HClO$_4$, and its anhydride, Cl$_2$O$_7$, have been known
for nearly two centuries\cite{Stadion1818}. Industrial uses of perchloric acid are manifold, and
higher chlorine oxides have recently been implicated in theories of stratospheric destruction of ozone\cite{envirostuff}. 
(See also two recent papers \cite{addref1,addref2} on the Cl$_2$O$_n$ and Br$_2$O$_n$ series (n=1--4) and
their anions.) Yet no reliable thermochemical data are available: the reason for this is probably best 
illustrated by the following quote from a spectroscopic study\cite{Wit73}:
\begin{quotation}
{\em Caution!} Several explosions occurred during the course of this work. It was necessary to perform all experiments wearing heavy gloves and a face shield with the sample properly shielded.
\end{quotation}

Cioslowski et al. compiled\cite{Cios2000} an extensive and chemically diverse set of thermochemical data for the purpose of benchmarking, calibration, and parametrization of computational thermochemistry methods. Their dataset includes Cl$_2$O$_7$, for which an unsourced value of $\Delta H^\circ_{f,298}$=65.0 kcal/mol was taken from the Wagman
et al. compilation\cite{Wag82}. An atom-equivalent scheme based on B3LYP/6-311++G** calculations proposed in Ref.\cite{Cios2000} yielded its single largest
error (91 kcal/mol) for this molecule. While it is to be expected that this
molecule will exhibit very strong inner polarization effects\cite{Bau95,so2},
it must also be said that the 65.0 kcal/mol number is a crude estimate, and
that any comparison with it will be semiquantitative at best.

For HClO$_4$, Francisco\cite{Fra95} quotes $\Delta H^\circ_{f,298}$=4$\pm$4 kcal/mol from the NIST database\cite{NIST1991}, and a group additivity estimate\cite{Colussi} of -1.5 kcal/mol. Francisco himself calculated $\Delta H^\circ_{f,0}$=+10.8 kcal/mol using G2 theory. Similar remarks apply as for Cl$_2$O$_7$. 
    
In the present work, we will apply W1 and W2 theory to these systems. We will also show that they are extreme cases of inner polarization effects\cite{Bau95,so2} and that
any calculation that does not adequately take this into account is doomed to failure.

\section{Computational Details}

All calculations were run on a 4-CPU AMD Opteron 846 server with 8 GB of RAM, two SATA system disks, and eight 72 GB UltraSCSI320 scratch disks, running SuSE Linux Enterprise Server 9 and custom-built for our group by Access Technologies of Re\d{h}ovot, Israel. The scratch disks are operated as two hardware RAID-0 arrays of four disks aggregated in software using the Linux ``md'' facility. For large soft-RAID "chunk sizes" of 512 KB and up, we are able to obtain sustained streaming read and write bandwidth in excess of 300 MB/s, as measured by IOzone\cite{iozone} for a 16 GB file.

The CCSD\cite{Pur82} and CCSD(T)\cite{Rag89,Wat93} coupled cluster calculations involved in W1 and W2 theory were carried out using MOLPRO 2002.6\cite{molpro} (always using conventional rather than integral-direct algorithms),
while all remaining calculations were carried out using a locally modified version of Gaussian 03\cite{g03}.
NBO (natural bond orbital)\cite{nbo} and AIM (atoms-in-molecules)\cite{aim} analyses were carried out using the relevant 
modules of Gaussian 03.

Three families of basis sets were used. The first are the original aug-cc-pVnZ (augmented correlation consistent polarized n-tuple zeta) basis sets of Dunning and coworkers\cite{avnz}: besides the unmodified 
aug-cc-pVnZ sequence, we considered the aug-cc-pVnZ+2d1f sequence proposed in the original W1/W2 paper\cite{w1}, where the``+2d1f'' notation represents the addition of two additional $d$ and one additional $f$ function, with exponents obtained by successively multiplying the highest exponent of that angular momentum already present by a factor of 2.5. 

The second family are the aug-cc-pV(n+d)Z basis sets of Wilson, Peterson, and Dunning\cite{PVX+dZ}, which were specifically developed to cope with inner polarization effects and contain an additional $d$ function beyond the original aug-cc-pVnZ basis sets.

The third family are the aug-pc$n$ (augmented polarization consistent) basis sets of Jensen\cite{Jensen1,Jensen2,Jensen3,Jensen4}. Unlike the correlation consistent basis sets which were developed for optimum recovery of the valence correlation energy, the polarization consistent basis sets were developed
for optimum recovery of Hartree-Fock and DFT energies, for which requirements (in terms of optimum exponents
and choice of polarization functions) are considerably different. We considered the aug-pc1, aug-pc2, aug-pc3, and aug-pc4 basis sets, which are of $3s2p$, $4s3p2d$, $6s5p3d2f$, and $8s7p4d3f2g$ quality, respectively, on hydrogen, of $4s3p2d$, $5s4p3d2f$, $7s6p5d3f2g$, and $9s8p7d4f3g2h$ quality, respectively, on oxygen, and of $5s4p2d$, $6s5p3d2f$, $7s6p5d3f2g$, and $8s7p7d4f3g2h$ quality, respectively, on chlorine. 
Here too, we considered addition of 
one or more high-exponent $d$ and $f$ functions (denoted by ``+d'', ``+2d'',\ldots suffixes), with exponents again
in even-tempered sequences with a stride factor of 2.5. 

Detailed descriptions (and rationales) for the different steps in W1 and W2 theory can be found in the original references\cite{w1,w1eval,w1review,w3}. In the interest of making the paper self-contained, we will briefly
summarize them:
\begin{itemize}
\item a B3LYP/cc-pV(T+d)Z\cite{B3LYP} reference geometry is used for W1 theory. Normally a CCSD(T)/cc-pV(Q+d)Z reference
geometry is called for in W2 theory: as full CCSD(T)/cc-pV(Q+d)Z optimizations proved computationally intractable, we have instead used the best available DFT geometries\cite{Boe05} (at the B97-1/aug-pc3+d level\cite{B971}).
The actual geometries used can be found in Figure 1; 
\item the SCF component is extrapolated from aug-cc-pVTZ+2d1f and aug-cc-pVQZ+2d1f basis sets in W1 theory,
and from aug-cc-pVQZ+2d1f and aug-cc-pV5Z+2d1f basis sets in W2 theory, using the formula\cite{w1eval} $E(L)=E_\infty+A/L^5$;
The acronyms W1w and W2w (rather than W1 and W2) indicate that aug-cc-pV(n+d)Z rather than aug-cc-pVnZ+2d1f basis sets were used\cite{w3};
\item the CCSD valence correlation component is extrapolated from the same pairs of basis sets using $E(L)=E_\infty+A/L^\alpha$, where $\alpha=3.22$ for W1 and W1w theory and $\alpha=3$ for W2 theory\cite{w1,l4,Hal98};
\item the (T) valence correlation component is extrapolated using the same expression, but from 
aug-cc-pVDZ+2d and aug-cc-pVTZ+2d1f basis sets in the case of W1 theory, and aug-cc-pVTZ+2d1f and aug-cc-pVQZ+2d1f basis sets for W2 theory (with the appropriate substitutions in the case of W1w and W2w). Note that the expensive (T) step need not be carried out in the largest basis set used overall. Note also that ROHF references are used for the open-shell atoms, and that the definition of Ref.\cite{Wat93} for the ROHF-CCSD(T) energy is used;
\item the differential contribution of inner-shell correlation is obtained at the CCSD(T) level with a specially developed MTsmall core correlation basis set\cite{hf,w1};
\item the scalar relativistic contribution can be obtained either as first-order mass-velocity and Darwin
corrections\cite{DMV} at the ACPF (averaged coupled pair functional\cite{Gda88}) level with the MTsmall basis set,
or (in the Gaussian 03\cite{g03} implementation of W1 theory and in the present work) as the difference 
between nonrelativistic and second-order Douglas-Kroll-Hess\cite{DKH} CCSD(T)/MTsmall energies;
\item These being closed-shell molecules, first-order spin-orbit corrections only affect the separated
atoms and are obtained from
the well-established atomic fine structures;
\item The zero-point vibrational energy was obtained from scaled B3LYP/cc-pV(T+d)Z harmonic frequencies 
for W1w theory, and from a recent DFT anharmonic force field study\cite{Boe05} for W2w theory;
\item Thermal corrections were obtained from the same data, although corrections are required for the hindered rotations in both molecules (see below).
\end{itemize}

\section{Results and discussion}

The reference geometries used are given in Figure 1.
Most salient results can be found in Table 1: basis set convergence for the SCF component
is presented in detail in Table 2. We shall discuss these results component by component.

\subsection{Hartree-Fock component}

Normally, the Hartree-Fock component is the easiest one to get right in an ab initio 
thermochemistry calculation. Yet in this case, very strong inner polarization effects
cause a sharp dependence of the Hartree-Fock component of the binding energy on the
presence of high-exponent d functions in the basis set. This can be seen in more detail
in Table 2.

For HClO$_4$ with the aug-cc-pVDZ or aug-pc1 basis sets, addition of four high-exponent $d$ functions (with exponents in an even-tempered series with stride 2.5) increases the computed binding energy by over 50 kcal/mol: for Cl$_2$O$_7$, the magnitude of the effect reaches a whopping 100 kcal/mol!

With aug-cc-pVTZ (or aug-pc2)  and $spdfg$ basis sets, the contributions 
for Cl$_2$O$_7$ go down to 45 and 23 kcal/mol, respectively --- the aug-pc3 basis set already contains one additional high-exponent d function to begin with, and thus requires less additional $d$ functions to reach saturation.

An NBO analysis of the wave function reveals that, while the chlorine $d$ orbitals have natural populations in the 0.35 range, these orbitals have Rydberg character and do not participate in any natural bond orbitals: as shown repeatedly previously\cite{CioslowskiMixon,ReedSchleyer}, there is no basis for describing HClO$_4$, nor by extension Cl$_2$O$_7$, as hypervalent molecules. 
NPA, APT (Atomic Polar Tensor\cite{APT}), and AIM all yield partial charges 
consistent with a general HO(Cl$^{+3}$)(--O$^-$)$_3$ bonding picture (see below).

What happens to the wavefunction when extra $d$ functions are added? 
Comparing NBO analyses for HClO$_4$ with aug-pc1 and aug-pc1+3d basis sets, we see that the energy of the 3d Rydberg orbital on Cl is lowered from
0.985 to 0.839 a.u. for $d_{xy}$, 0.974 to 0.830 a.u. for $d_{xz}$, 1.200 to 1.016 a.u. for $d_{yz}$, 1.134 to 0.966 a.u. for $d_{x^2-y^2}$,
and 1.173 to 0.996 a.u. for $d_{3z^2-r^2}$. Thus the orbitals become more accessible: overall $3d$ population increases from 0.28 to 0.37
electron equivalents, with the corresponding NBO occupations clustering in a "quasi-$t_{2g}$" triad of 0.117, 0.103, and 0.100 and a "quasi-$e_g$" dyad of 0.0594 and 0.0588. (These numbers add up to more than 0.37 as there are small non-$d$ components in these orbitals.) Interestingly, the
Wiberg bond indices for the three short Cl--O bonds go up by about 0.05, and of the long bond by 0.02. The increase in the AIM (Bader) covalent bond orders is more pronounced: from 1.27 to 1.42--1.43 for the short bonds, and from 1.00 to 1.05 for the long bond.
Second-order perturbation theory analysis of the NBOs reveals, besides the expected interactions between oxygen lone pairs and antibonding ClO orbitals, strong interactions between oxygen lone pairs and the $3d$ Rydberg orbitals on chlorine.

Consistent with all of this, the AIM charge distribution is considerably less polarized with versus without the extra $d$ functions ($\delta$(Cl)=3.51 and 3.92, respectively). The NPA charges are not significantly affected ($\delta$(Cl)=2.68 in both cases), while the APT charges actually show a mild opposite trend 
(2.58 vs. 2.53).

Summing up, the chemical significance of the extra $d$ functions is thus that they improve the ability of the $3d$ orbital to act as an acceptor for backbonding from the oxygen lone pair orbitals. This situation is, in fact, somewhat reminescent of the role of the 3d orbital in
the CaO molecule\cite{fritzCaO,group12}. 

As a result of all this, basis set convergence for the aug-cc-pVnZ series is quite unsatisfactory. Somewhat more satisfying results are obtained with the Wilson-Peterson-Dunning aug-cc-pV(n+d)Z series, although even here, there is still an unusually large increment
of 0.68 kcal/mol (for Cl$_2$O$_7$: 1.29 kcal/mol)
between the aug-cc-pV(5+d)Z and aug-cc-pV(6+d)Z basis sets. With the aug-cc-pV$n$Z+2d1f series advocated in the original
W1/W2 paper\cite{w1}, convergence is rather more satisfactory: the aug-cc-pV5Z+2d1f basis set actually yields a larger binding energy (78.81 kcal/mol) than the aug-cc-pV(6+d)Z and aug-cc-pV6Z+d basis set (78.70 and 78.76 kcal/mol, respectively). Our best directly computed value, with the aug-cc-pV6Z+2d1f
basis set, is only 0.05 kcal/mol greater. Note that the contribution of the high-exponent $f$ function decays rapidly in the aug-cc-pVnZ+2d1f 
series: from 0.98 kcal/mol for $n$=T to 0.02 kcal/mol for $n$=6. Note also that even for the aug-cc-pV6Z+d basis set, adding a second hard $d$ function still contributes 0.08 kcal/mol.

Turning to Jensen's polarization consistent basis sets --- which were specifically developed for SCF or DFT applications rather than for correlated ab initio calculations --- we still see a 0.58 kcal/mol increment between uncontracted aug-pc3+d and aug-pc4 basis sets. (Jensen recommends using uncontracted basis sets for SCF energy extrapolations.) Using Jensen's recommended extrapolation formula\cite{Jensen2} from aug-pc2+2d, aug-pc3+d, and aug-pc4 data (see footnote c to Table 1 for details), we obtain a best 
estimate for TAE(SCF)=78.94 kcal/mol: for comparison, the best directly computed values are 78.70 kcal/mol for aug-cc-pV(6+d)Z, 
78.84 kcal/mol for aug-cc-pV6Z+2d,
78.86 kcal/mol for aug-cc-pV6Z+2d1f,
and 
78.77 kcal/mol for uncontracted aug-pc4. 

Using the aug-cc-pV\{5,6\}Z+2d1f data instead with the $A+B/L^5$ extrapolation
used in W2 theory, we obtain 78.89 kcal/mol as our limit: taking into account Jensen's observation that the SCF energy for such large basis sets appears to converge much faster than $L^5$ (see also Ref.\cite{Schwenke05}), we take the average of this extrapolation and
the raw aug-cc-pV6Z+2d1f value, to finally obtain TAE(SCF)=78.88 kcal/mol.
The W1w extrapolation falls far short at 76.35 kcal/mol, while the performance of the W2w
extrapolation, 78.61 kcal/mol, could be deemed acceptable. (With the "+2d1f" basis set series of standard W2 theory, we would obtain 79.09 kcal/mol, which 
overshoots the true limit.) Using the efficient direct SCF codes presently available, 
it might well be that the 
preferred approach for the SCF component in problematic cases like this
--- rather than an extrapolation on purely empirical grounds --- would be to carry out a
single straight SCF calculation with the largest routinely feasible basis set. Even in the present extreme case, using the raw
SCF/aug-cc-pV6Z+2d1f or SCF/aug-pc4+d numbers causes errors under 0.1 kcal/mol.

An SCF/aug-cc-pV6Z+2d1f calculation on Cl$_2$O$_7$, which requires no less than 1,743 basis functions, 
yields TAE(SCF)=-30.81 kcal/mol. $A+B/L^5$ extrapolation yields -30.76 kcal/mol, averaging as for HClO$_4$ -30.79
kcal/mol. (The raw SCF/aug-cc-pV(6+d)Z and SCF/aug-pc4 values are in error by 0.31 and 0.19 kcal/mol, respectively,
suggesting aug-cc-pV6Z+2d1f to be the basis set of choice for "single point" SCF limit calculations.)
In terms of convergence along basis set sequences, the same trends as for HClO$_4$ are seen in amplified form. We note in passing that the SCF reaction energy of the isodesmic reaction 2~${\rm HClO}_4$ $\rightarrow$ ${\rm Cl}_2{\rm O}_7$ + ${\rm H}_2{\rm O}$ converges quite rapidly with the basis, as long as at least one hard $d$ function is present.

\subsection{Valence correlation energy}

As noted previously for SO$_2$ and SO$_3$,\cite{so2,so3} basis set convergence for the valence correlation
component is not anomalous for molecules with severe inner polarization. As for any molecule with highly polar bonds, however
(see, e.g., SiF$_4$ \cite{sif4}, BF$_3$ \cite{bf3}, and the like), a significant difference between W1w and W2w is to be expected. 
In the case of HClO$_4$, the discrepancy for the CCSD correlation energy is quite modest at 0.5 kcal/mol, and it is partly compensated
by a discrepancy of -0.2 kcal/mol for the connected triple excitations contribution. 

Even a CCSD/aug-cc-pV(Q+d)Z calculation on Cl$_2$O$_7$ was only barely feasible with the available hardware: CCSD(T)/aug-cc-pV(Q+d)Z and
CCSD/aug-cc-pV(5+d)Z calculations are plainly impossible. Considering the good agreement between W1 and W2 for the valence correlation
contributions, however, we can safely assume that this component is well reproduced for the isodesmic reaction.

Neither molecule shows any obvious sign of severe static correlation, neither from the ${\cal T_1}$ diagnostics\cite{T1diag} (0.019 for HClO$_4$, 0.021 for Cl$_2$O$_7$), nor from the largest coupled
cluster amplitudes. Yet the fact that three-quarters of the atomization energy of HClO$_4$ results from correlation, and  that
Cl$_2$O$_7$ is actually slightly metastable at the Hartree-Fock level, suggest that higher-order correlation effects could be somewhat
important for these molecules. Unfortunately, a W3 calculation for HClO$_4$ --- let alone Cl$_2$O$_7$ --- is absolutely impossible with
the present state of technology. This appears to be the single greatest source of uncertainty in our calculations.

\subsection{Inner-shell correlation energy}

Like with other second-row molecules, despite the rather large absolute inner-shell correlation energies, the inner-shell contribution
to the molecular binding energy is rather small (0.90 kcal/mol for HClO$_4$). A CCSD(T)/MTsmall calculation on Cl$_2$O$_7$ is not
feasible. We were able to carry out a CCSD(T)/cc-pwCVTZ calculation, but this basis set is clearly woefully inadequate for HClO$_4$, 
recovering only about half the contribution. We might be able to rely on error compensation for the isodesmic reaction, and 
the inner-shell correlation contribution to its reaction energy is found to be -0.18 kcal/mol at the CCSD(T)/cc-pwCVTZ level.
From the higher-level results for HClO$_4$ and H$_2$O, we can then extract a best estimate of 1.38 kcal/mol for the inner-shell contribution
to the Cl$_2$O$_7$ binding energy.

\subsection{Relativistic effects}

The original W1 and W2 protocols called for mass-velocity and Darwin corrections from ACPF/MTsmall calculations. W3 theory, as well as 
the W1 and W2 implementations in the popular Gaussian 03 program package, use CCSD(T) calculations with the Douglas-Kroll approximation
to obtain the scalar relativistc contribution.
We have followed the same (more rigorous) approach here, as the scalar relativistic contributions are generally significant for this
type of molecule (second-row atoms in high oxidation states surrounded by strongly electronegative elements). At the DK-CCSD(T)/MTsmall
level, we find -2.69 kcal/mol 
for HClO$_4$ and a hefty -4.80 kcal/mol for Cl$_2$O$_7$. For HClO$_4$, we recalculated the contribution using relativistically optimized
correlation consistent basis sets\cite{w3} of AVTZ and AVQZ quality. We obtain fundamentally the same result, and upon extrapolation
to the infinite-basis limit, we actually reproduce the MTsmall result to two decimal places. We therefore have not attempted any DK-CCSD(T)/AVQZ calculation for Cl$_2$O$_7$: for the sake of completeness, the DK-CCSD(T)/AVTZ value
is -4.73 kcal/mol.

The first-order spin-orbit contribution results exclusively from the atomic fine structures and requires no further comment. We expect second-order spin-orbit coupling corrections to be on the order of 0.1 kcal/mol or less, well below the more important potential error sources in these calculations.

\subsection{Zero-point and thermal corrections}

The scaled B3LYP/cc-pV(T+d)Z harmonic frequencies normally used for the zero-point energy in W1w theory is clearly falling short a bit in these cases. From a very recent DFT anharmonic force field study\cite{Boe05} on both molecules, we have available anharmonic ZPVEs at the B97-1/aug-pc3+d level. Our "best estimate" value for HClO$_4$ was obtained by combining the anharmonic ZPVE with one-half the difference between the computed and observed fundamental frequencies. For Cl$_2$O$_7$ the assignment of the experimental vibrational spectrum is too fraught with
ambiguities to allow for the same approach: here we have assumed the ratio between the "best estimate" and directly computed ZPVEs for HClO$_4$ to be transferable to Cl$_2$O$_7$. 

The thermal correction represents the minor complication of hindered ClO$_3$ rotations. In the case of HClO$_4$, we find the internal rotation
barrier $V$ to be 0.647 kcal/mol at the B97-1/aug-pc2+2d level, which corresponds to a $V/RT$ ratio of nearly one. At the same level,
we obtain a reduced moment for the internal rotor of 2.9853 amu.bohr$^2$. With a de-facto rotor periodicity of three, we obtain $1/Q_f=0.52$ from eq.(4) of Pitzer and Gwinn\cite{PitzerGwinn}. By interpolating their Table V, we obtain 1.054 e.u., or 0.53 R, for $E_{introt}/T$, i.e., very close to
the free rotor limit of $R/2$ and considerably removed from the low-frequency harmonic vibration limit of $R$. Our `best estimate' values reflect this correction for HClO$_4$, and twice this correction for Cl$_2$O$_7$.
Clearly, it will cancel in the isodesmic reaction.

\subsection{Heats of formation}

Our final best estimate for the heat of formation of HClO$_4$ is -0.6 kcal/mol. We expect this value to be accurate to about 1 kcal/mol, with the largest potential source of error being post-CCSD(T) correlation
effects. 

This value actually agrees quite well with Colussi and Grela's group additivity estimate of -1.5 kcal/mol.
The large discrepancy between the G2 theory\cite{g2} result of Francisco\cite{Fra95},
$\Delta H^\circ_{f,0}=$+10.8 kcal/mol and our own
absolute-zero heat of formation of 2.4 kcal/mol should not be surprising considering the
extreme basis set sensitivity pointed out here. Using the more up-to-date G3 theory\cite{g3}, we obtain
TAE$_0$=307.9 kcal/mol (compared to 305.4 kcal/mol for G2), 
or 306.2 kcal/mol after applying the atomic spin-orbit correction (-1.73 kcal/mol): 
G3 thus still underbinds the molecule by 7.6 kcal/mol compared to our best estimate
(TAE$_0$=313.8 kcal/mol). G3X theory\cite{g3x}, which was recently developed in an attempt to (inter alia)
reduce errors for (pseudo)hypervalent compounds, recovers an additional 5.64 kcal/mol worth of binding energy and thus reduces
the error to a respectable 2 kcal/mol. While G3 uses an MP2/6-31G* reference geometry (which will have 
very substantial basis set incompleteness error for this type of molecules), G3X uses B3LYP/6-31G(2df,2p)
--- in particular the extra $d$ function very significantly affects the computed reference geometry.
In addition, G3X involves a step where a single $g$ function is added to the basis set at the HF level:
this does affect the energy by 2.01 kcal/mol, which appears to be a slight overestimate, as we find the
$g$ function contribution in the aug-pc3 basis set to be only 1.45 kcal/mol.

Our best estimated heat of formation for Cl$_2$O$_7$ (obtained from the W1w isodesmic reaction energy and 
our best calculated data for HClO$_4$ and H$_2$O) is 65.9 kcal/mol, to which we attach a conservative error bar of 2 kcal/mol. This is actually quite close to the Wagman et al.\cite{Wag82} estimate of $\Delta H^\circ_{f,298}$=65.0 kcal/mol and the R\"uhl et al.\cite{Ruh99} mass spectrometric value of 65$\pm$4 kcal/mol. 
Li et al.\cite{Li2000} used a variety of methods from the G2 and G3 family, and found 
76.8 kcal/mol at the G3 level, 84.9 kcal/mol at the G3(MP3) level, and 66.7 kcal/mol at the G2(MP2) level, which is fortuitously in much better agreement with our best value than the more sophisticated G3-based methods.
Once again, we find that G3X theory puts in a much better performance: after spin-orbit correction,
we obtain TAE$_0$=399.47 kcal/mol, again within 2 kcal/mol of our best estimate. The G3X $g$ function is found
to account for 3.94 kcal/mol (once again, an overestimate compared to the 2.86 kcal/mol from the $g$ functions
in the aug-pc3 basis set), with most of the rest once again being accounted for by the superior reference
geometry.

Sicre and Cobos\cite{Sic03} obtained 93.1 kcal/mol at the B3LYP/6-311+G(3d2f) level, 78.9 kcal/mol at the
mPW1PW91 level with the same basis set, 86.2 kcal/mol at the G3(MP2)//B3LYP level and 79.5 kcal/mol at
the G3(MP2)//B3LYP/6-311+G(3d2f) level. These same authors' best estimate, from isodesmic reaction schemes,
is 61.5 kcal/mol, appreciably below our best estimate.

\section{Conclusions}

The heats of formation of HClO$_4$ and Cl$_2$O$_7$ have 
been determined to chemical
accuracy 
for the first time 
by means of W1 and W2 theory. These molecules exhibit particularly severe degrees of inner polarization, and as such obtaining a basis-set limit SCF component to the total atomization energy 
becomes a challenge. (Adding high-exponent $d$ functions to a standard $spd$ basis set has an effect
on the order of 100 kcal/mol for Cl$_2$O$_7$.) Wilson's aug-cc-pV(n+d)Z basis sets represent a dramatic
improvement over the standard aug-cc-pVnZ basis sets, while the aug-cc-pVnZ+2d1f sequence converges still
more rapidly. Jensen's polarization consistent basis sets still require additional high-exponent $d$
functions: for smooth convergence we suggest the \{aug-pc1+3d,aug-pc2+2d,aug-pc3+d,aug-pc4\} sequence.
The role of the tight $d$ functions is shown to be an improved description of the Cl (3d) Rydberg orbital, enhancing its ability to receive back-bonding from the oxygen lone pairs. In problematic cases like this (or indeed in general),
a single 
SCF/aug-cc-pV6Z+2d1f calculation
may be preferable over empirically motivated extrapolations.
Our best estimate heats of
formation are $\Delta H^\circ_{f,298}[$HClO$_4$(g)$]=-0.6\pm$1 kcal/mol and 
$\Delta H^\circ_{f,298}[$Cl$_2$O$_7$(g)$]=65.9\pm$2 kcal/mol, the largest source of uncertainty being our 
inability to account for post-CCSD(T) correlation effects. While G2 and G3 theory have fairly large errors,
G3X theory reproduces both values to within 2 kcal/mol.

\section{Acknowledgments}
This work was
supported by the Lise
Meitner-Minerva Center for Computational Quantum Chemistry 
(of which JMLM is a member {\em ad personam}) 
and by the Helen and Martin Kimmel Center for
Molecular Design. It is related to Project 2003-024-1-100, "Selected Free Radicals and Critical Intermediates: Thermodynamic Properties from Theory and Experiment," of the International Union of Pure and Applied Chemistry (IUPAC).

\clearpage

\begin{table}
\caption{Contributions to the total atomization energies and heats of formation of HClO$_4$ and Cl$_2$O$_7$. All data in kcal/mol}
\begin{tabular}{lcccccccc}
\\
\hline\hline
                 &\multicolumn{3}{c}{HClO$_4$}&\multicolumn{2}{c}{Cl$_2$O$_7$}&\multicolumn{2}{c}{H$_2$O}&(a)\\                                    &   W1w  & W2w     &best    &   W1   &   best  & W1   &   W2     & W1w\\
\hline
SCF limit        &   76.35& 78.61   &78.88(c)& -32.46 &  -30.79 &159.90& 160.01   &+25.26     \\
CCSD-SCF limit   &  228.59&229.10   &229.10  &  399.33&         & 69.62&  69.20   &-11.77     \\ 
(T) limit        &   27.07& 26.87   &26.87   &  54.61 &         &  3.64&   3.55   & -4.11     \\
inner-shell corr.&    0.90&  0.96   & 0.96   & [1.44](b)& 1.38(d)  &  0.36&   0.36   & +0.18 (d) \\
scalar rel.      &   -2.66& -2.69   &-2.69(i)& -4.80  &         & -0.26&  -0.26   & -0.26     \\
spin-orbit       &   -1.73& -1.73   &-1.73   &  -3.24 &         & -0.22&  -0.22   &  0.00     \\
TAE$_e$          &  328.51&331.12   &331.39  &  414.88&420.84(k)&233.03& 232.64   & +9.30     \\
ZPVE             &   17.07& 17.44(f)&17.64(g)&  19.16 & 19.52(h)& 13.15&  13.15   & -1.83     \\
TAE$_0$          &  311.44&313.68   &313.75  &  395.72&401.32(k)&219.88& 219.49   & +7.47     \\
$\Delta H^\circ_{f,0}$&4.72& 2.48   & 2.41   &  74.36 & 69.53(k)&-57.63& -57.24   & +7.47     \\
$H_{298}-H_0$    &    3.56&  3.56   & 3.28(j)&   5.81 &  5.25(j)&  2.37&   2.37   & +1.62     \\
$\Delta H^\circ_{f,298}$&2.03&-0.21 &-0.57   &  70.71 & 65.87(k)&-58.32& -57.92(e)& +9.09     \\
\hline\hline
\end{tabular}

\begin{flushleft}{\scriptsize

(a) reaction energy of 2 HClO$_4$ $\rightarrow$ Cl$_2$O$_7$ + H$_2$O

(b) Best estimate 1.44 kcal/mol if isodesmic reaction assumed to be thermoneutral w.r.t. inner shell correlation

(c) HClO$_4$: From $A+B/L^5$ and AV\{5,6\}Z+2d1f data: 78.89 kcal/mol. Assuming this overestimate, and taking average of raw value and extrapolated, best estimate
is 78.88 kcal/mol. Cl$_2$O$_7$: -30.76 and -30.79 kcal/mol, respectively.
basis sets and Jensen's extrapolation formula
$E(L,n_s)=E_\infty+A(L+1)\exp(-B \sqrt{n_s})$,  with $n_s$ taken for Cl: 
B=4.901885, $E_\infty$=78.94 kcal/mol.
(Nonlinear equation solved using "goalseek" feature of Excel.)

(d) From CCSD(T)/cc-pwCVTZ calculations. With cc-pwCVTZ basis set: 0.336 kcal/mol H2O, 0.581 kcal/mol Cl$_2$O$_7$, 0.546 HClO$_4$, hence $\Rightarrow$ -0.175 for reaction. Hence best estimate for Cl$_2$O$_7$ from reaction and best available data for H$_2$O and HClO$_4$: 1.38 kcal/mol

(e) Expt. value: -57.80 kcal/mol

(f) From B97-1/aug-pc3+d anharmonic force field\cite{Boe05}

(g) From anharmonic force field and $\sum_i{\nu_i({\rm expt.})-\nu_i({\rm calc.})}$=0.20 kcal/mol 

\clearpage

(h) From anharmonic force field\cite{Boe05}: ZPVE=19.30 kcal/mol. Expt. spectral assignment problematic\cite{Boe05}: assuming similar relative error as for HClO$_4$ yields scaling factor of 1.0115, hence best estim. ZPVE=19.52 kcal/mol

(i) extrapolated from -2.64 and -2.67 kcal/mol, respectively, with relativistic correlation consistent basis sets (see Ref.\cite{w3})

(j) from Gaussian calc. with hindered rotor stuff: reduced moment=3.0142 amu.bohr$^2$. In these units,
eq. (4) of Pitzer \& Gwinn\cite{PitzerGwinn} becomes $Q_f=2.815\sqrt{0.00465 I_{red}T}/n$, with rotor 
periodicity $n$ de facto =3. Hence $1/Q_f$=0.52; with $V/RT\approx1$, from Table V of Pitzer \& Gwinn we obtain by interpolation $E_{\rm int.rot.}=0.53 R$. Thus the RRHO $H_{298}-H_0$ has to be reduced by 0.47 RT=0.28 kcal/mol at 298 K.

(k) from W1 reaction energy of (a) and best calculated data for HClO$_4$ and H$_2$O

}\end{flushleft}

\end{table}
\clearpage
\begin{table}
\caption{Basis set convergence of the SCF contribution to the total atomization energies (kcal/mol) of 
HClO$_4$ and Cl$_2$O$_7$}
\begin{tabular}{lrrrrrrr}
\hline\hline
            & unmodified&+d & +2d   & +3d   & +4d   & (X+d)Z& +2d1f\\
\hline
\multicolumn{8}{c}{HClO$_4$}\\
\hline
aug-cc-pVDZ & -1.77 & 36.31 & 48.90 & 51.77 & 52.17 & 37.18 & 48.90\\
aug-cc-pVTZ & 52.35 & 68.20 & 74.47 & 75.21 & 75.33 & 73.01 & 75.45\\
aug-cc-pVQZ & 65.03 & 75.29 & 78.15 & 78.47 & 78.50 & 76.82 & 78.24\\
aug-cc-pV5Z & 75.56 & 78.39 & 78.74 &       &       & 78.02 & 78.81\\
aug-cc-pV6Z & 77.77 & 78.76 & 78.84 &       &       & 78.70 & 78.86\\
aug-pc1uncon&   0.55&       &       & 53.56 & 53.99 &       & 53.99\\
aug-pc2uncon&  66.24&       & 71.41 & 71.42 &       &       & 71.98\\
aug-pc3uncon&  78.01& 78.19 & 78.25 &       &       &       & 78.24\\
aug-pc4uncon&  78.77& 78.78 &       &       &       &       & 78.78\\
aug-pc1 & 0.90\\
aug-pc2 & 67.47\\
aug-pc3 & 78.82\\
aug-pc4 & 79.45\\
\hline
\multicolumn{8}{c}{Cl$_2$O$_7$}\\
\hline
aug-cc-pVDZ & -182.36 & -109.78 & -85.15 & -79.69 & -78.95 & -106.60 & -85.15\\
aug-cc-pVTZ & -81.41 & -50.85 & -38.68 & -37.25 & -37.03 & -41.49 & -36.89\\
aug-cc-pVQZ & -57.45 &  &  &  &  & -34.60 & -31.89\\
aug-cc-pV5Z & -37.17 &  &  &  &  & -32.41 & -30.90\\
aug-cc-pV6Z & -32.91 &  &  &  &  & -31.12 & -30.81\\
aug-pc1uncon& -178.19\\
aug-pc2uncon& -55.04\\
aug-pc3uncon& -32.46\\
aug-pc4uncon& -30.98\\
\hline\hline
\end{tabular}

\end{table}
\clearpage
\begin{figure}
\includegraphics[width=12cm,angle=0]{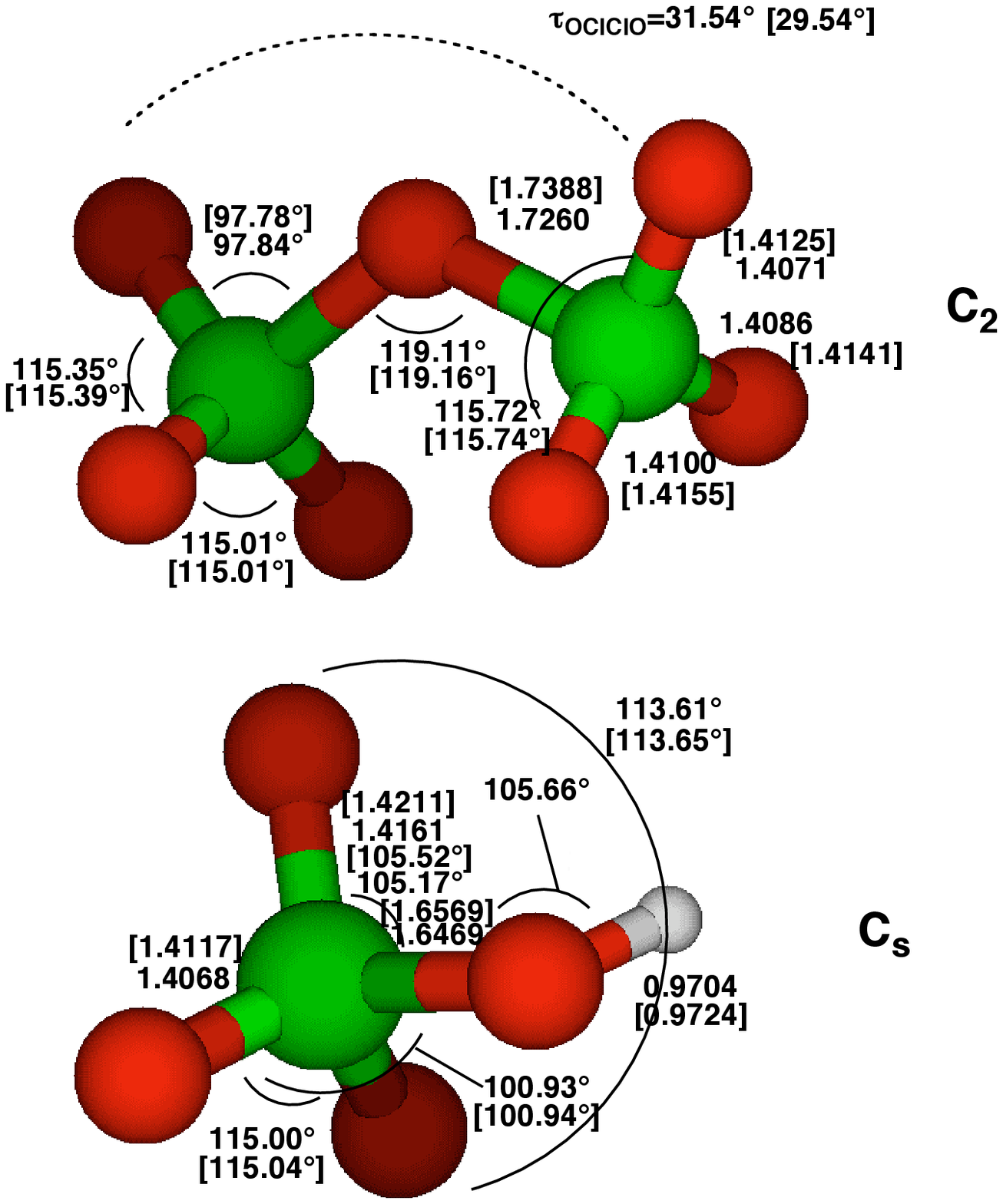} \\
\caption{B3LYP/cc-pV(T+d)Z (in square brackets) and B97-1/aug-pc3+d geometries\cite{Boe05} for HClO$_4$ and Cl$_2$O$_7$(\AA, degrees)}
\end{figure}

\end{document}